\documentclass[11pt,a4paper]{article}
\pdfoutput=1
\usepackage{jcappub}
\usepackage{subfig}

\newcommand{\ARdelete}[1]{}

\newcommand{\nc}{\newcommand}
\nc{\bea}[1]{\begin{eqnarray} \mbox{$\label{#1}$}}
\nc{\eea}{\end{eqnarray}}
\nc{\be}[1]{\begin{equation} \mbox{$\label{#1}$}}
\nc{\ee}{\vspace{0.1cm}\end{equation}}
\nc{\eq}[1]{\mbox{Eq.\ (\ref{#1})}}
\nc{\fig}[1]{\mbox{Fig.\ \ref{#1}}}
\nc{\ch}[1]{\mbox{Chapter\ \ref{#1}}}
\nc{\sect}[1]{\mbox{Section\ \ref{#1}}}
\nc{\HRule}{\rule{\linewidth}{0.5mm}}
\nc{\nl}{\nonumber \\}

\nc{\hf}{\frac{1}{2}}
\nc{\hfrt}{\frac{1}{\sqrt{2}}}
\nc{\fb}[2]{\left(\frac{#1}{#2}\right)}
\nc{\sqb}[2]{\sqrt{\frac{#1}{#2}}}

\nc{\fnl}{f_{\rm NL}}

\def\GeV{{\rm \ GeV}}

\begin{document}
\setlength{\unitlength}{1mm}
\title{Electrically charged curvaton}

\author[a,b]{Michela D'Onofrio,}
\author[a,b]{Rose N. Lerner}
\author[c]{and Arttu Rajantie}

\affiliation[a]{University of Helsinki, P.O. Box 64, FI-00014, Helsinki, Finland.}
\affiliation[b]{Helsinki Institute of Physics, P.O. Box 64, FI-00014, Helsinki, Finland.}
\affiliation[c]{Department of Physics, Imperial College London,
London SW7 2AZ, United Kingdom}

\emailAdd{michela.donofrio@helsinki.fi}
\emailAdd{rose.lerner@helsinki.fi}
\emailAdd{a.rajantie@imperial.ac.uk}


\abstract{

We consider the possibility that the primordial curvature perturbation was generated through the curvaton mechanism from a scalar field with an electric charge, or precisely the Standard Model U(1) weak hypercharge. This links the dynamics of the very early universe concretely to the Standard Model of particle physics, and because the coupling strength is known, it reduces the number of free parameters in the curvaton model. The gauge coupling also introduces several new physical effects.
Charge fluctuations are generated during inflation, but they are screened by electron-positron pairs therefore do not violate observational constraints. After inflation, the curvaton interacts with thermal radiation which destroys the curvaton condensate and prevents the generation of curvature perturbations, unless the inflaton dynamics satisfy strong constraints. The curvaton also experiences a period of parametric resonance with the U(1) gauge field. Using the standard perturbative approach, we find that the model can generate the observed density perturbation for Hubble rate $H_* \gtrsim 10^8 \GeV$ and curvaton mass $m \gtrsim 10^{-2}H_*$, but with a level of non-Gaussianity ($f_{NL} \gtrsim 130$) that violates observational constraints. However, previous studies have shown that the parametric resonance changes the predicted perturbations significantly, and therefore fully non-linear numerical field theory simulations are required.
}

\maketitle

\section{Introduction}

The inflationary paradigm is commonly accepted as the origin of the primordial fluctuations that seeded structure in the Universe. Models of inflation are usually based on quantum field theory, with fields unrelated to the Standard Model of particle physics. However, some interaction between the inflaton and the Standard Model fields is required in order to reheat the Universe, which means that a single theoretical framework for both inflation and particle physics is ultimately required.

If the inflaton field is responsible for generating the primordial curvature perturbations, its coupling to other fields has to be extremely weak in order to avoid radiative corrections that would spoil the flatness of the potential. This makes it difficult to couple the inflaton to Standard Model gauge fields, which all interact relatively strongly. A less constrained alternative is the curvaton model \cite{Linde:1996gt,Enqvist:2001zp,Lyth:2001nq,Moroi:2001ct}, in which the perturbations are generated by a separate scalar field known as the curvaton. The curvaton is light and subdominant during inflation and gains an isocurvature perturbation. After inflation has ended, the energy density of the curvaton grows relative to the background radiation. When the curvaton finally decays, its isocurvature perturbation is converted to an adiabatic perturbation that can seed the structure in the Universe.

The purpose of this paper is to investigate whether the curvaton field could be charged under a Standard Model gauge group. This would have the attractive feature that the properties of these interactions are known, in contrast with typical curvaton models, which have so much freedom that it is hard to make definite predictions. Of the three Standard Model gauge groups, the most promising is the U(1) weak hypercharge, because the SU(2) and SU(3) groups are confining and would give more complicated physics. In contrast, a U(1) charge would essentially mean that the curvaton has an electric charge, and it is relatively easy to investigate the various constraints that arise from this.

To be specific, we consider the Lagrangian
\be{lag}
{\cal L} =  - m^2 \sigma^\dagger \sigma - \lambda (\sigma^\dagger \sigma)^2- \frac{1}{4}F_{\mu\nu}F^{\mu\nu} + \left| (i\partial_\mu - g' A_\mu)\sigma \right|^2,
\ee
where $\sigma$ is the curvaton field, $g'$ is the Standard Model U(1) gauge coupling, $A_\mu$ is the Standard Model U(1) gauge field and $F_{\mu\nu}$ is the corresponding field strength tensor. The coupling strength is known~\cite{pdg}, $g' \approx 0.36$, and thus the number of free parameters is reduced.\footnote{The coupling will run to larger values --- we quote the weak scale value.}
For concreteness, we assume that the curvaton carries one unit of hypercharge,
$Y=1$. The hypercharge has to be an integer to allow the curvaton to decay into Standard Model particles, and a higher value would not change our conclusions significantly. For $Y=2$, the curvaton could have a Yukawa coupling to right-handed electrons, but for other values the only other renormalisable term is a bilinear coupling $\sigma^\dagger\sigma \Phi^\dagger\Phi$ to the Higgs field. For simplicity, we assume that this coupling is negligible. The opposite case would essentially correspond to the model discussed in Refs.~\cite{Enqvist:2008be,Chambers:2009ki}.

To determine whether such a model is viable, we apply both theoretical and observational constraints, for example requiring a stable vacuum and the correct amplitude of the curvature perturbation $\zeta$. In this model, three mechanisms affect the curvaton's dynamics after inflation: non-perturbative parametric resonance into photons, interactions with the thermal bath, and perturbative decay. We find that a perturbative estimate of the non-Gaussianity ($f_{\rm NL}$) appears to rule out the model at 95\% confidence level, but this ignores the parametric resonance which is known to have a significant effect on perturbations~\cite{Chambers:2009ki} and therefore it cannot be relied on. Instead, non-linear field theory simulations are ultimately needed to determine the perturbations.

In \sect{over}, we give a brief overview of dynamics of the model.  In \sect{charge}, we discuss the charge fluctuations produced during inflation and show that, because of electric screening, they are compatible with observations. Then in \sect{curvaton}, we give the conditions that the model must satisfy to have the qualitative behaviour of a curvaton model. These are conditions such as requiring a light curvaton during inflation and no false vacuum in the curvaton potential. In \sect{decay}, we then expand the discussion of the dynamics after inflation, analytically calculating timescales for each process. In \sect{results} we discuss the generation of curvature perturbations and production of transient cosmic string-like structures. We conclude in \sect{conclusions}.

\section{Overview of the model}
\label{over}

The electrically charged curvaton model follows the standard curvaton scenario in many ways, but with more complicated dynamics and tighter constraints. We first review the standard curvaton scenario. There are two fields: the inflaton and the curvaton. During inflation, the inflaton dominates the energy density, but its perturbations are assumed to be negligible. The curvaton field is light (compared to the Hubble rate\footnote{We use subscript $*$ to represent values at the end of inflation.} $H_*$), and therefore it develops a nearly scale-invariant spectrum of fluctuations, in the same way as the inflaton. After inflation, it is assumed that the inflaton reheats to produce a thermal bath, which becomes the dominant form of energy in the Universe.

As the Universe expands, $H$ decreases, and when $H\lesssim m$, the curvaton begins to oscillate in its potential.
Assuming a harmonic potential,\footnote{Other potentials have been discussed in Refs.~\cite{Enqvist:2009zf,Enqvist:2009ww}.} its equation of motion is
\be{eomSlin}
\ddot{\sigma} + 3 H \dot{\sigma} + m^2  \sigma =0.
\ee
In a radiation-dominated background, $H(t) = 1/(2 t)$ and $a(t)= \sqrt{2H_*t}$.
The evolution of $\sigma$ is then given by
\bea{sigmaexactsolution}
\sigma(t) &=& \frac{2^{1/4} \ \Gamma\left(\frac{5}{4}\right) \sigma_*}{(mt)^{1/4}} \ J_{1/4}\left(m t\right) \nl
 &\approx& 0.86
\frac{\sigma_*}{\left(m t \right)^{3/4}} \ \cos (mt - 3\pi/8).  
\eea
The energy density of the curvaton field,
\be{rho_res}
\rho_{\sigma} \approx m^2\sigma(t)^2\approx  0.74 \frac{m^{1/2}\sigma_*^2}{t^{3/2}},
\ee
decreases as $a^{-3}$,
whereas that of the radiation,
\be{rho_gamma}
\rho_\gamma = \frac{3M_p^2 H_*^2}{a^4} = \frac{3M_p^2}{4 t^2}.
\ee
decreases proportional to $a^{-4}$.
Thus, the relative fraction of curvaton $r(t)$ grows,
\be{r_dec}
r(t) \equiv \frac{3\rho_\sigma(t)}{3\rho_\sigma(t) +
4\rho_\gamma(t)}\propto t^{1/2} . 
\ee

Eventually, the curvaton decays to Standard Model particles which have the equation of state of radiation --- at this point the perturbations in the curvaton become adiabatic. Assuming that this happens instantaneously at time $t_{\rm dec}$, the amplitude of the curvature
perturbation is given by \cite{Lyth:2001nq}
\be{zeta0}
\zeta \approx \frac{r(t_{\rm dec})}{3}\frac{\delta\rho_\sigma}{\rho_\sigma},
\ee
This needs to agree with the observed amplitude $\zeta \simeq 10^{-5}$.
In the standard curvaton model, in which Eq.~(\ref{rho_res}) remains valid until the decay time, this becomes
\be{zeta}
\zeta \approx \frac{r(t_{\rm dec})}{3}\frac{2}{\sigma_*}\delta\sigma_*\approx \frac{H_* r(t_{\rm dec})}{3\pi \sigma_*}.
\ee
With the same assumption, the non-Gaussianity of the perturbations is given by
\be{fnl}
f_{\rm NL} \approx \frac{5}{4r_{\rm dec}}.
\ee

In the electrically charged curvaton model, the dynamics between the end of inflation and $\zeta$ becoming adiabatic are non-trivial. The large coupling $g'$ means that photon-curvaton interactions play an important role, through a number of processes. A thermal bath containing photons will interact with the condensate. This thermal bath could either be produced at the end of inflation by reheating of the inflaton, or later by curvaton decay products. The fate of the model strongly depends on whether any interactions with the thermal bath lead to full chemical equilibrium. If they do, then $\zeta$ becomes adiabatic when that happens. Otherwise, the relative fraction of curvaton can continue to grow until the final decay.

As we will see in Section~\ref{sec:nonpert}, the large photon-curvaton coupling also leads to resonant production of photons as the condensate oscillates. This changes $\rho_\sigma$, and makes it a more complicated function of $\sigma_*$. Therefore Eqs.~(\ref{zeta}) and (\ref{fnl}) are no longer valid. If the resonance ends before backreaction makes the dynamics non-linear, some condensate will remain, and in principle the dependence can be calculated from linear theory.
Previous numerical simulations indicate that some non-relativistic particles will
remain~\cite{Chambers:2009ki} even if the dynamics become non-linear, but in that case a full lattice field theory simulation~\cite{Chambers:2007se,Chambers:2008gu} is required to calculate the dependence of $\rho_\sigma$ on $\sigma_*$. This result could then be subsituted to Eq.~(\ref{zeta0}) to compute the curvature perturbation.
Simulations of this type are, however, beyond the scope of this paper, and therefore we will use Eq.~(\ref{zeta}) to obtain a rough estimate of the perturbations.

We also assume that there is some mechanism for a perturbative decay of the curvaton. This is likely to be suppressed because there are no direct 3-point couplings to the Standard Model. There is, however, a strict lower bound on the curvaton's final effective decay rate ($\Gamma~\gtrsim~10^{-22}\GeV$) from requiring the curvaton to decay before BBN. To avoid large isocurvature perturbations, the curvaton should also decay before dark matter freezes out, but this constraint is strongly model-dependent{\footnote{For reference, WIMP dark matter that decouples at $T\sim {\cal O}(10)\GeV$ gives $\Gamma \gtrsim 10^{-15}\GeV$.}} and we do not impose it. Both the resonance and thermal interactions are likely to occur too early to produce sufficient~$\zeta$. However, provided that \emph{some} non-relativistic curvaton remains after these processes, the curvaton's energy density can continue to grow, before perturbatively decaying and producing the majority of the curvature perturbation.

Either the interactions of an initial thermal bath with the condensate, or the resonant production of photons by the condensate will leave a distribution of curvaton particles that may not be in chemical equilibrium.{\footnote{We are careful to distinguish between kinetic equilibrium (distribution of momenta) and chemical equilibrium (number densities of particles).}} At this point, the evolution depends on the temperature. If $T\ll m$, then the curvaton's energy density continues to evolve as $\rho_\sigma \propto a^{-3}$, and a late perturbative decay will eventually give an adiabatic $\zeta$, which could be large enough to match observations. If $T \gg m$ but chemical equilibrium is never reached, then the curvaton energy density scales like $\rho_\sigma \propto a^{-4}$ until the temperature drops and the curvatons freeze out. From this point, the evolution is the same as the $T\ll m$ case, and a large $\zeta$ is possible. Alternatively, if chemical equilibrium \emph{is} reached, then the
number density of the curvatons is in equilibrium, and $\zeta$ becomes adiabatic. In this case, $\zeta$ is unaffected by the freeze-out or decay of the curvatons, and is likely to be very small (because chemical equilibrium occurs at an early time).

\section{Field and Charge Fluctuations}
\label{charge}

\subsection{Field fluctuations}
\label{fieldfluct}
The curvaton is a light field, and will therefore gain perturbations during inflation, in the same way as the inflaton field.
Assuming for simplicity that $m=0$, the linearised equation
of motion for a mode with comoving momentum $\mathbf{k}$ is
\be{eom0}
\ddot\sigma(\mathbf{k})+3H\dot{\sigma}(\mathbf{k})+\frac{k^2}{a^2}\sigma(\mathbf{k})=0,
\ee
where $k=|\mathbf{k}|$.
The mode is initially in its vacuum state, and when it crosses the horizon, i.e., $k/a=H$, it freezes out,  leading to a scale-invariant spectrum of fluctuations
\be{sigmacorr}
\langle \sigma^\dagger(\mathbf{k}) \sigma(\mathbf{q})\rangle=(2\pi)^3\delta(\mathbf{k}+\mathbf{q})G_{00}(k),
\ee
with
\be{G_00}
G_{00}(k)=\frac{H^2}{2k^3}.
\ee

For comparison with observations, we split the field into a background part $\sigma_*$ which is the average field value in our currently observable universe, and fluctuations $\delta\sigma_*$ which are local deviations from this background value. That is, within the observable universe we have
\be{sigmasplit}
\sigma(\mathbf{x})=\sigma_*+\delta\sigma_*(\mathbf{x}).
\ee
The precise value of $\sigma_*$ is not determined by the theory, but its typical value can be estimated by calculating the variance
\be{variance}
\langle |\sigma_*|^2 \rangle=\int_{\Lambda_{\rm IR}}^{\Lambda_{\rm CMB}} \frac{d^3k}{(2\pi)^2}G_{00}(k)
=\frac{H^2}{4\pi^2}\ln\frac{\Lambda_{\rm CMB}}{\Lambda_{\rm IR}}
,
\ee
where $\Lambda_{\rm CMB}$ is
the comoving horizon size at the time when the largest observable scales left the horizon,
and $\Lambda_{\rm IR}$ is the comoving horizon size at the start of inflation.
This can also be expressed as
\be{varianceN}
\langle |\sigma_*|^2 \rangle
=\left(N_{\rm tot}-N_{\rm CMB}\right)\frac{H_*^2}{4\pi^2},
\ee
where $N_{\rm tot}$ is the total number of e-foldings of inflation, and $N_{\rm CMB}\approx 60$ is the number of
e-foldings after the largest observable scales left the horizon. Although $N_{\rm tot}$ is not determined by the theory, we would typically expect $N_{\rm tot}-N_{\rm CMB}\gg 1$, and therefore $\sigma_*\gtrsim H_*$. When the non-zero mass of the curvaton is taken into account, this estimate eventually breaks down when the probability distribution function for $\sigma_*$ approaches the equilibrium solution, with $\langle |\sigma_*|^2 \rangle \approx \frac{3}{4\pi^2} \fb{H_*}{m}^2 H_*^2$ \cite{Starobinsky:1994bd}. For large $m/H_*$, the approach to this equilibrium solution is particularly fast \cite{Enqvist:2012xn}. The initial conditions for the curvaton model are discussed further in \cite{Enqvist:2012xn}.

In the same way, the variance of the field fluctuations $\delta\sigma_*$ relative to the background value is
\be{variancedelta}
\langle |\delta\sigma_*|^2 \rangle=\int_{\Lambda_{\rm CMB}}^{H_*} \frac{d^3k}{(2\pi)^2}G_{00}(k)
=\frac{H_*^2}{4\pi^2}\ln\frac{H_*}{\Lambda_{\rm CMB}}=N_{\rm CMB}\frac{H_*^2}{4\pi^2}.
\ee

\subsection{Charge fluctuations}

One obvious question about the electrically charged curvaton is whether it generates a non-zero charge density with observable consequences, either during or after inflation. To address this, we have to calculate the charge density and charge distribution of the curvaton field. Because the electric charge is conserved, any charge density on large scales survives at least until the relevant scale enters the horizon. Therefore, there would still be a non-zero electric charge density in the observable Universe due to the fluctuations of the curvaton field, and we have to check if it is compatible with observations.
The current bound on a homogeneous electric charge density per baryon is
$q_{e-p} < 10^{-26}e$ \cite{Caprini:2003gz}. If the charge distribution is not symmetric, this bound relaxes to $q_{e-p} < 5\times 10^{-20}e$ \cite{Caprini:2003gz}.

A large field value does not necessarily imply high charge density.
Instead, the charge density per physical volume of a complex scalar field $\sigma$ is given by
\be{chargedensity0}
Q=g'{\rm Im}\,\left(\dot\sigma^*\sigma\right).
\ee
Because this involves the time derivative of the fields, it is suppressed on superhorizon scales where the field evolution is described the by overdamped equation
\be{eom}
3H\dot{\sigma}(\mathbf{k})+\frac{k^2}{a^2}\sigma(\mathbf{k})=0.
\ee

To calculate the statistical properties of the charge density, we write the complex curvaton field $\sigma_*$
field in terms of its real and imaginary parts $\sigma=(\sigma_1+i\sigma_2)/\sqrt{2}$, which both have the same
correlation function
\be{compcorr}
\langle \sigma_1(\mathbf{k}) \sigma_1(\mathbf{q})\rangle=\langle \sigma_2(\mathbf{k}) \sigma_2(\mathbf{q})\rangle=(2\pi)^3\delta(\mathbf{k}-\mathbf{q})G_{00}(k).
\ee
In terms of these fields, the charge density is
\be{chargedensity}
Q=\frac{g'}{2}\left(\dot{\sigma}_1\sigma_2-\sigma_1\dot{\sigma}_2\right).
\ee
We can use Eq.~(\ref{eom}) to calculate the correlation functions for the time derivatives,
\bea{compcorrderiv}
\langle \sigma_1(\mathbf{k}) \dot\sigma_1(\mathbf{q})\rangle&=&\langle \sigma_2(\mathbf{k}) \dot\sigma_2(\mathbf{q})\rangle=(2\pi)^3\delta(\mathbf{k}-\mathbf{q})G_{01}(k),
\nonumber\\
\langle \dot\sigma_1(\mathbf{k}) \dot\sigma_1(\mathbf{q})\rangle&=&\langle \dot\sigma_2(\mathbf{k}) \dot\sigma_2(\mathbf{q})\rangle=(2\pi)^3\delta(\mathbf{k}-\mathbf{q})G_{11}(k),
\eea
and we find
\be{G_xx}
G_{01}(k) = -\frac{H}{6a^2k} \;\;\;\;{\mbox {and}}\;\;\;\; G_{11}(k) = \frac{k}{18a^4}.
\ee

On average the charge density is zero because fluctuations of either sign are equally likely. However, there will be local charge fluctuations with the two-point correlation function
\be{chargecorr}
\langle Q(\mathbf{x})Q(\mathbf{y})\rangle = g'^2\left[G_{11}(|\mathbf{x}-\mathbf{y}|)G_{00}(|\mathbf{x}-\mathbf{y}|)-G_{01}(|\mathbf{x}-\mathbf{y}|)^2\right].
\ee
Transforming \eq{chargecorr} to momentum space, we obtain
\begin{eqnarray} \label{twopoint}
\langle Q(\mathbf{k})Q(\mathbf{q})\rangle &=&
g'^2(2\pi)^3\delta(\mathbf{k}+\mathbf{q}) \int \frac{d^3p}{(2\pi)^3}\bigl[
G_{11}(\mathbf{p})G_{00}(\mathbf{k}-\mathbf{p})-G_{01}(\mathbf{p})G_{01}(\mathbf{k}-\mathbf{p})\bigr]\nl
&=&\frac{g'^2H^2}{72a^4}
(2\pi)^3\delta(\mathbf{k}+\mathbf{q}) \int \frac{d^3p}{(2\pi)^3}
\frac{\bigl((\mathbf{k}-\mathbf{p})^2-p^2\bigr)^2}{p^3(\mathbf{k}-\mathbf{p})^3}.
\end{eqnarray}
We can approximate this integral by noting that the dominant contribution comes from the two singularities $\mathbf{p}=0$ and $\mathbf{p}=\mathbf{k}$. The integrand is symmetric, so the two poles are identical. Thus, expanding around $\mathbf{p}=0$, we find
\begin{eqnarray}
\langle Q(\mathbf{k})Q(\mathbf{q})\rangle
&\approx&\frac{g'^2H^2}{36a^4}
(2\pi)^3\delta(\mathbf{k}+\mathbf{q}) \int \frac{d^3p}{(2\pi)^3}
\frac{k}{p^3},
\end{eqnarray}
The UV divergence is cut off by $k$, and assuming an infrared cutoff $\Lambda\ll k$, we have
\begin{eqnarray}
\langle Q(\mathbf{k})Q(\mathbf{q})\rangle
\approx \frac{g'^2H^2k}{72\pi^2a^4}
(2\pi)^3\delta(\mathbf{k}+\mathbf{q}) \int_{\Lambda}^k \frac{dp}{p}
&= &(2\pi)^3\delta(\mathbf{k}+\mathbf{q}) \frac{g'^2H^2}{72\pi^2a^4}k\log\frac{k}{\Lambda}\nl
&=& (2\pi)^3\delta(\mathbf{k}+\mathbf{q}) G_{Q}(k),
\end{eqnarray}
where we have defined
\begin{equation}
\label{equ:Grho}
G_{Q}(k)=\frac{g'^2H^2}{72\pi^2a^4}k\log\frac{k}{\Lambda}.
\end{equation}
This correlation function describes the charge density on superhorizon scales.
By Fourier transforming it, we can write down the correlation function in coordinate space
\be{chargecorr_coord}
\langle Q(0)Q(\mathbf{x})\rangle \approx
\int\frac{d^3k}{(2\pi)^3}e^{i\mathbf{k}\cdot\mathbf{x}}G_{Q}(k)=\frac{g'^2H^5}{432\pi^4}\ln\left(\frac{aH}{\Lambda}\right)\frac{1}{ar},
\ee
where $r$ is comoving distance. This shows that there are long-range power-law charge fluctuations whose amplitude is growing as the logarithm of the scale factor.

In order to see whether the electromagnetic interactions have a significant effect on the field dynamics, we estimate the typical electric field $\mathbf{E}$ on the scale of the horizon. From Gauss's law we find that, in momentum space, the electric field due to the charge distribution is given by
\be{Efield}
\mathbf{E}(\mathbf{k})=\frac{ia\mathbf{k}Q(\mathbf{k})}{k^2},
\ee
and therefore the variance of the electric field on the horizon scale is
\be{Evar}
\langle \mathbf{E}^2\rangle
=\int \frac{d^3k}{(2\pi)^3}
\frac{a^2}{k^2}G_{Q}(k)
\approx \frac{{g'}^2H^4}{288\pi^4}\ln\frac{aH}{\Lambda}.
\ee
To check whether this field influences the dynamics significantly, we compare it to the acceleration due to the vacuum energy during inflation, which is
\be{accelH}
\ddot{\vec{x}}=H^2\vec{x}.
\ee
Estimating $\ln(aH/\Lambda)\sim 100$, the typical acceleration due to the electric field,
\be{accelE}
|\ddot{\vec{x}}|=\frac{g'}{m}|\vec{E}|\sim 0.01\frac{H^2}{m},
\ee
is significant at distances $|\vec{x}|\lesssim 0.01/m$. In this model, we find that $10^{-2}H_* \lesssim m \lesssim H_*$ (see \sect{results}), so this effect could indeed be relevant on super-horizon scales.

However, as long as $H>m_e/{g'}$, where $m_e$ is the electron mass, the electric field (\ref{Evar}) is strong enough to create electron-positron pairs through the Schwinger process~\cite{Schwinger:1951nm}. Therefore the charge fluctuations of the curvaton field will be screened by an opposite fluctuation in the electron-positron charge density. The remaining electric fields $|\vec{E}|\sim m_e^2/g'$ are far too weak to give rise to significant acceleration, and therefore we conclude that the backreaction to the curvaton evolution is likely to be negligible, meaning that the curvaton fluctuations have the usual nearly scale-invariant spectrum. For the remainder of the paper, this effect is ignored. As a result of the screening, the Universe will also be charge neutral, and therefore there is no conflict with the observational bounds on the charge density of the Universe.

\section{The Effective Potential}
\label{curvaton}

The large value of the curvaton gauge coupling $g'$ gives rise to corrections to the potential, which can have a substantial impact on the parameter space of the model \cite{Enqvist:2011jf}. Assuming that the one-loop corrections are dominated by the gauge coupling $g'$, the effective potential is \cite{Coleman:1973jx}
\be{CWterm}
V_{\rm eff}(\sigma) = m^2 |\sigma|^2  +   \frac{3g'^4}{64\pi^2}|\sigma|^4 \ln\left(\frac{|\sigma|^2}{\mu^2}\right) ,
\ee
where we have absorbed the self-coupling constant $\lambda$ into the definition of the renormalisation scale $\mu$. Therefore, the free parameters are $m$ and $\mu$.

\begin{figure}[b]
\begin{centering}
\vspace{0.5cm}
 \includegraphics[width=8cm]{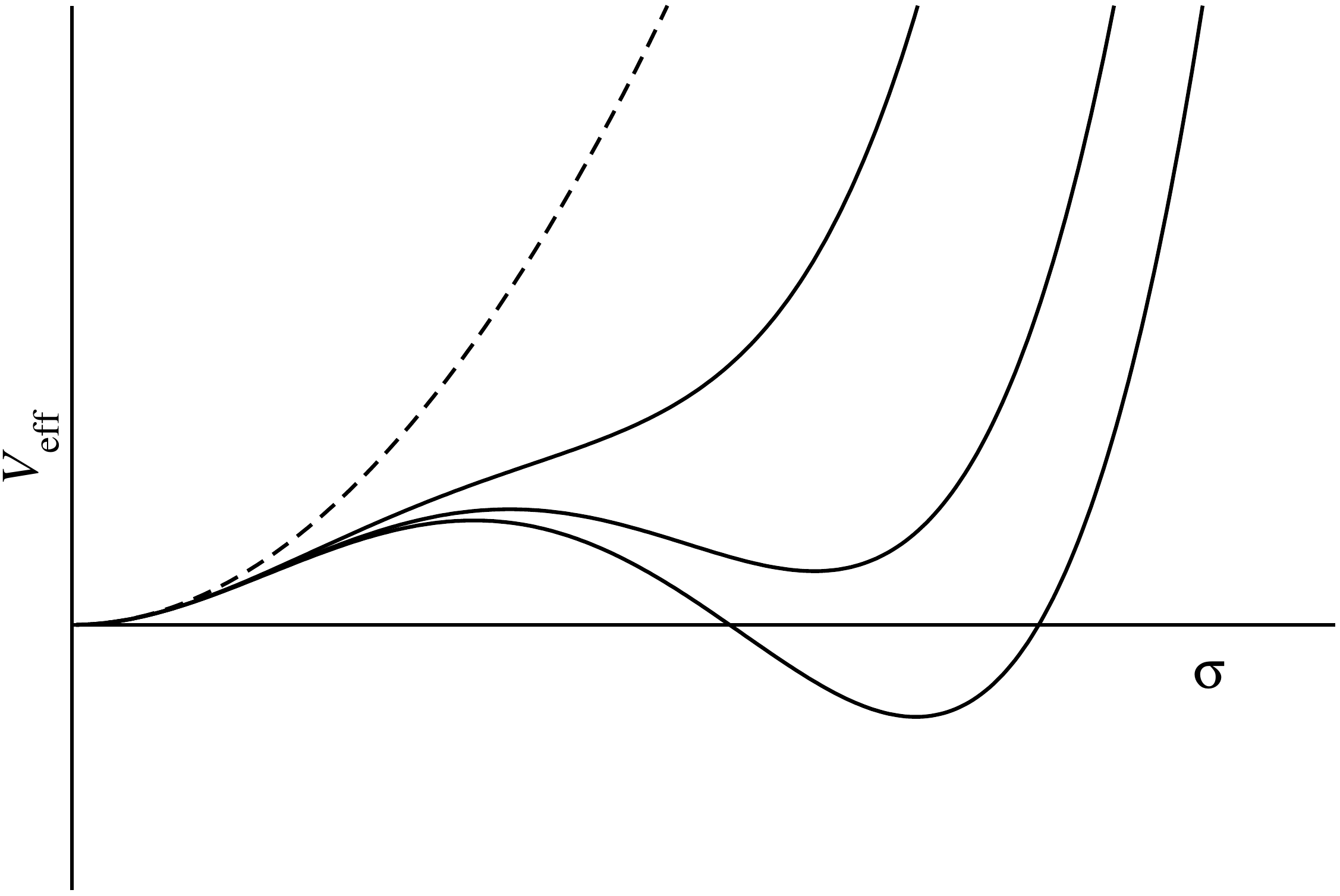}
\caption{\footnotesize{The effective potential $V_{\rm eff}(|\sigma|)$ of the curvaton field. The dashed line shows the quadratic tree-level potential, and the three solid lines show the effective potential for increasing $\mu$, from left to right.}}
\label{fig:effpot}
\end{centering}
\end{figure}

The shape of this potential is shown in Fig.~\ref{fig:effpot}.
For small $\mu$, the U(1) symmetric vacuum $\sigma=0$ is the only minimum. At larger $\mu$, a second minimum with $\sigma\ne 0$ appears, and when $\mu$ is large enough,
this symmetry-breaking minimum becomes the true vacuum.

In order for $\sigma$ to act as a curvaton field, the effective potential has to satisfy certain conditions:
\begin{enumerate}
 \item {\bf Vacuum stability:} The symmetric vacuum $\sigma=0$ has to be the true vacuum. Otherwise the Universe would tunnel into the vacuum with $\sigma\ne 0$, which would break the U(1) symmetry spontaneously and make the photon massive. This is not allowed by the experimental constraints on the photon mass and gives the constraint
\begin{equation}
\mu^2<\frac{64\pi^2\exp(1)}{3g'^4}m^2.
\end{equation}
\item {\bf Shallow potential:}
In order to gain a nearly scale-invariant spectrum of perturbations during inflation, the curvaton field has to be light compared to the Hubble rate $H_*$.
This means that its effective mass $m_{\rm eff}$, defined as the second derivative of the effective potential at the relevant field value, has to be less than the Hubble rate during inflation, so
\be{m_req}
m_{\rm eff}^2 \equiv V''_{\rm eff}(\sigma_*)\ll H_*^2.
\ee
In addition, the curvaton should be subdominant compared with the inflaton potential because otherwise it is an inflaton. This means that the effective potential has to satisfy
\be{sd_req}
V_{\rm eff}(\sigma_*) \ll 3 M_p^2 H_*^2.
\ee
Both of these conditions, (\ref{m_req}) and (\ref{sd_req}) restrict the height of the effective potential, and also the maximum value of the curvaton field $\sigma_*$.
\item {\bf Linearity:}
The curvaton has to evolve linearly both during and after inflation. During inflation, we have to make sure that
the effective mass (\ref{m_req}) is light over the whole range of field values present in the observable universe.
This imposes the constraint
\be{lin1_req}
\delta\sigma_*V_{\rm eff}'''(\sigma_*)\ll H_*^2,
\ee
where, according to Eq.~(\ref{variancedelta}), $\delta\sigma_*\sim \sqrt{N}H_*$.

After inflation, we require that the potential is dominated by the mass term $m^2|\sigma|^2$, so
\be{quad_req}
m^2\gg \left|
\frac{3g'^4}{64\pi^2}\sigma_*^2 \ln\left(\frac{\sigma_*^2}{\mu^2}\right)\right|.
\ee
This implies that the curvaton has the equation of state of matter after inflation, making it easy to study the dynamics. The condition also guarantees that if a metastable second minimum exists, the field is on the left side of the barrier and starts to oscillate around the symmetric minimum.
We note that it may be possible to relax the constraint (\ref{quad_req}), and have a curvaton model where the potential is initially dominated by the correction term. This has been discussed in Ref.~\cite{Enqvist:2011jf}.
\end{enumerate}

\begin{figure}[!tb]
\begin{centering}
\includegraphics[width=0.5\textwidth, angle=270]{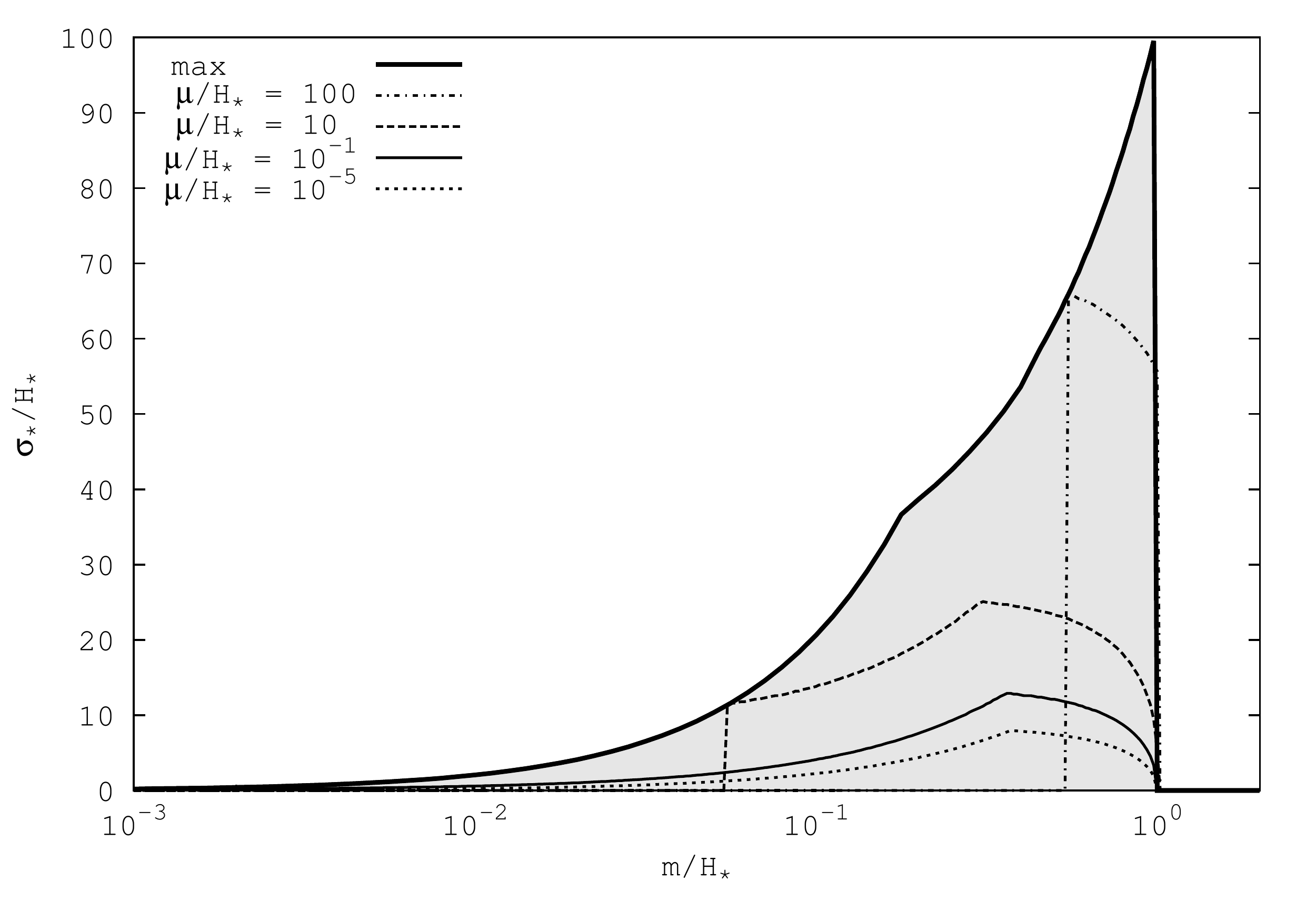}
\caption{\footnotesize{Allowed region (shaded) for a viable curvaton model satisfying conditions 1.--3. in the text. Note that these conditions only guarantee the existence of a curvaton, but not that the generated perturbations are compatible with observations. Also shown are the maximum $\sigma_*$ for various constant $\mu$; vacuum instability rules out large $\mu$ and small $m$. The parameter space is independent of $H_*$. This is because $H_*$ only determines the subdominance of the curvaton during inflation, which is naturally satisfied provided the other conditions are met.}
\label{pspace_mu}}
\end{centering}
\end{figure}

The effect of these constraints is shown in \fig{pspace_mu} (allowed region is shaded). They are highly non-trivial, both on the value of $\sigma_*$, and on the mass of the field (in generic curvaton models, the curvaton mass $m$ can be as low as $10^{-10}H_*$). The combination of these constraints favours large curvaton mass $m \gtrsim 10^{-2} H_*$, relatively (but not unnaturally) low field values $\sigma_* \lesssim {\cal O}(100) H_*$, and $\mu \lesssim {\cal O}(100) H_*$. The maximum $\sigma_*$ for various constant values of $\mu$ is also shown, to demonstrate that larger values of $\sigma_*$ are only possible for larger values of $\mu$.

\section{Evolution after inflation}
\label{decay}

After the end of inflation, the curvaton field consists of a homogeneous \emph{condensate} that oscillates in its potential according to Eq.~(\ref{sigmaexactsolution}). The evolution of this condensate is affected by its interactions with other fields, which cause it to decay into curvaton \emph{particles} and other degrees of freedom. By curvaton particles, we mean any curvaton field modes with non-zero momentum, and for simplicity we assume that they have a thermal spectrum, so that their behaviour is completely parameterised by their number density and temperature.

The U(1) charge of the curvaton field does not allow a direct Yukawa coupling to Standard Model fermions, but we assume that there is some indirect decay channel, through non-renormalisable interactions or some beyond-the-Standard-Model fields. This allows the curvaton to decay perturbatively into fermions at a slow rate $\Gamma$ which we treat as a free parameter. At earlier times the curvaton's interactions are dominated by its U(1) gauge coupling, which has two effects: it allows the curvaton condensate to decay non-perturbatively to photons through a parametric resonance; and thermal photons (if present) interact with the curvaton condensate, turning it into curvaton particles.

These processes affect the behaviour and energy density of the curvaton field and therefore the curvature perturbation $\zeta$ through Eq.~(\ref{zeta}). The curvature perturbation becomes adiabatic when the curvaton either decays or obtains full (chemical and kinetic) thermal equilibrium with the dominant background. To obtain a sufficiently high amplitude of perturbations, it is usually necessary for the curvature perturbation $\zeta$ to become adiabatic very late. Note that the mechanism of decay can also affect the non-Gaussianity of the model.

From \fig{pspace_mu} we can see that $\sigma_*\lesssim 100H_*$. Together with \eq{zeta}, this implies that in order to generate the observed amplitude of perturbations the energy fraction in the curvaton field $r_{\rm dec}$ must less than $10^{-2}$. Therefore we can safely assume that the curvaton's contribution to the energy density is subdominant and the background energy density scales like radiation.  As discussed in
Section~\ref{fieldfluct}, field values well below the Hubble rate are unnatural, and therefore we also set a lower limit $\sigma_* > 0.1H_*$.

We now discuss these processes in detail, beginning with the effect of a thermal bath created by the inflaton.

\subsection{Interactions with the thermal bath}

Because the curvaton-photon coupling $g'$ is relatively large, the interaction of the curvaton with any thermal bath containing photons is significant. In particular, interactions with photons kick curvaton particles out of the condensate, creating a thermal spectrum of curvaton particles. This process has the rate~\cite{Thoma:1996ag}
\be{Gamma_thermal}
\Gamma_{\rm th} \approx 0.03 g'^2 T,
\ee
where $T$ is the temperature of the thermal bath.

This thermal interaction only converts the condensate into curvaton particles, but does not cause the curvaton particles to decay. However, the non-perturbative decay of the curvaton into photons (through parametric resonance) is then prevented, because this process requires a homogeneous condensate. If the temperature is low ($T\ll m$), then the produced particles are non-relativistic and the equation of state is unchanged compared to that of the oscillating condensate. If the temperature is high enough ($T\gg m$), the produced curvaton particles are ultrarelativistic and have the equation of state of radiation. In this case, the evolution depends on whether both chemical and kinetic equilibrium are reached.

In our scenario, a thermal bath of radiation can be generated immediately after inflation by the decaying inflaton field, or later by annihilating curvaton particles. Let us first focus on the former case. Assuming that the inflaton has decayed completely to thermal radiation, the temperature of this radiation is given by (assuming $g_* = 100$)
\be{temp}
T \simeq 0.5 \sqrt{M_p H}.
\ee
According to \eq{Gamma_thermal}, this gives an effective thermalisation rate
\be{Gamma_th_inf}
\Gamma_{\rm th} \simeq 10^{-3} \sqb{M_p}{H} H.
\ee
The process is efficient if the thermalisation timescale is less than the age of the Universe i.e., $1/\Gamma_{\rm th} \lesssim 1/H$.
For all reasonable values of parameters, thermalisation occurs effectively instantaneously (with $T\gg m$), destroying the curvaton condensate. If the curvaton also reaches chemical equilibrium, this makes the produced
curvature perturbation $\zeta$ too small to be compatible with observations.

Chemical equilibrium requires annihilation or pair creation of curvaton-anticurvaton pairs.
We estimate the rate of these processes as
\bea{Gamma_chem}
\Gamma_{\rm chem} & \approx& \frac{g'^4}{16\pi^2} T_{\rm th}
\eea
which is suppressed relative to \eq{Gamma_th_inf} by two powers of the coupling $g'$. Chemical equilibrium is achieved if $1/\Gamma_{\rm chem} \ll 1/H$, which is quickly satisfied for all reasonable values of parameters, compared to the timescale needed to produce sufficiently large $\zeta$. For large $H_*$ and $m$, the timescale may be comparable to that of the non-perturbative effects, discussed below.

Thus, in order to have a viable model we must impose the requirement that there are (almost) no photons in the thermal background after inflation. This could occur either (i) if the inflaton decays to a hidden sector, sufficiently decoupled from the Standard Model, or (ii) if the inflaton is blocked from decaying until a sufficiently late time.{\footnote{Even a small fraction of the inflaton decaying into Standard Model particles could be enough to cause thermalisation of the curvaton condensate. The exact limits depend on the parameters.}} In this second case, we require the inflaton to oscillate in a $\phi^4$ potential in order that $r$ can grow sufficiently. We assume that one of these two possibilities can occur.

\subsection{Non-perturbative decay of the curvaton}
\label{sec:nonpert}
  Provided we avoid thermalisation by the inflaton's decay products, the first interaction of the curvaton is non-perturbative production of photons, through a parametric resonance~\cite{Traschen:1990sw,Kofman:1994rk,Rajantie:2000fd}. This is particularly important because of the large photon-curvaton coupling. Note, however, that charge conservation prevents perturbative decay through this coupling.

\ARdelete{
From the Lagrangian (\eq{lag}), the full equation of motion for the curvaton is
\begin{equation}\label{eomSfull}
\ddot{\sigma} + 3 H \dot{\sigma} - \left( \frac{ie\partial_i A_i}{a^2} - \frac{g'^2 A^2}{a^2} - m^2 \right) \sigma = 0.
\end{equation}
In our analytical calculations, we use the linear approximation \ref{eomSlin}, which is valid because $A_\mu$ is in the vacuum state during inflation and therefore ${ie\partial_i A_i}/{a^2} \ll m^2$.
}
In the temporal Coulomb gauge, the equation of motion for the gauge field $\mathbf{A}$ in the homogeneous background (\ref{sigmaexactsolution}) is
\be{eomA}
\mathbf{\ddot{A}}(t,\mathbf{k}) + H(t) \mathbf{\dot{A}}(t,\mathbf{k}) + \left( \frac{k^2}{a(t)^2}  + 2 g'^2 \sigma(t)^2\right) \mathbf{A}(t,\mathbf{k}) = 0,
\ee
where $\mathbf{k}$ is the comoving momentum.
The gauge field has been normalised in such a way that for $g'=0$ it oscillates with a constant amplitude.

Rescaling the gauge field $\mathbf{A}(t,\mathbf{k})$ with 
$ \mathbf{B}(t,\mathbf{k}) = a(t)^{1/2} \mathbf{A}(t,\mathbf{k})$ and substituting for $\sigma$ using (\eq{sigmaexactsolution})
gives a Mathieu equation with time-dependent parameters,
\be{Mathieu}
\mathbf{B}^{''}(z,\mathbf{k}) + \left( \Sigma_k(z) + 2q(z) \cos 2 z \right) \mathbf{B}(z,\mathbf{k}) = 0,
\ee
where $z = mt-3\pi/8$ is a dimensionless time parameter and $'$ denotes differentiation with respect to $z$. The parameters are
\bea{coeffs1}
q(z)     = \frac{2^{1/2}\Gamma(5/4)^2}{\pi}\frac{g'^2\sigma_*^2}{m^2}\left(z+\frac{3\pi}{8}\right)^{-3/2}
\approx
0.37 \frac{  g'^2  \sigma_*^2}{\ m^2  z^{3/2}} 
\eea
and
\bea{coeffs2}
\Sigma_k(z) &=& \frac{k^2}{a^2 m^2} + \frac{1}{4 m^2} \left( \frac{\dot{a}}{a}\right)^2 - \frac{1}{2 m^2} \left(\frac{\ddot{a}}{a}\right)  + 2q(z) \nonumber\\
&=&\frac{k^2}{2mH_*(z+3\pi/8)}+\frac{3}{16(z+3\pi/8)^2}+2q(z).
\eea

\begin{figure}[!ht]
\begin{center}
\includegraphics[width=0.7\textwidth]{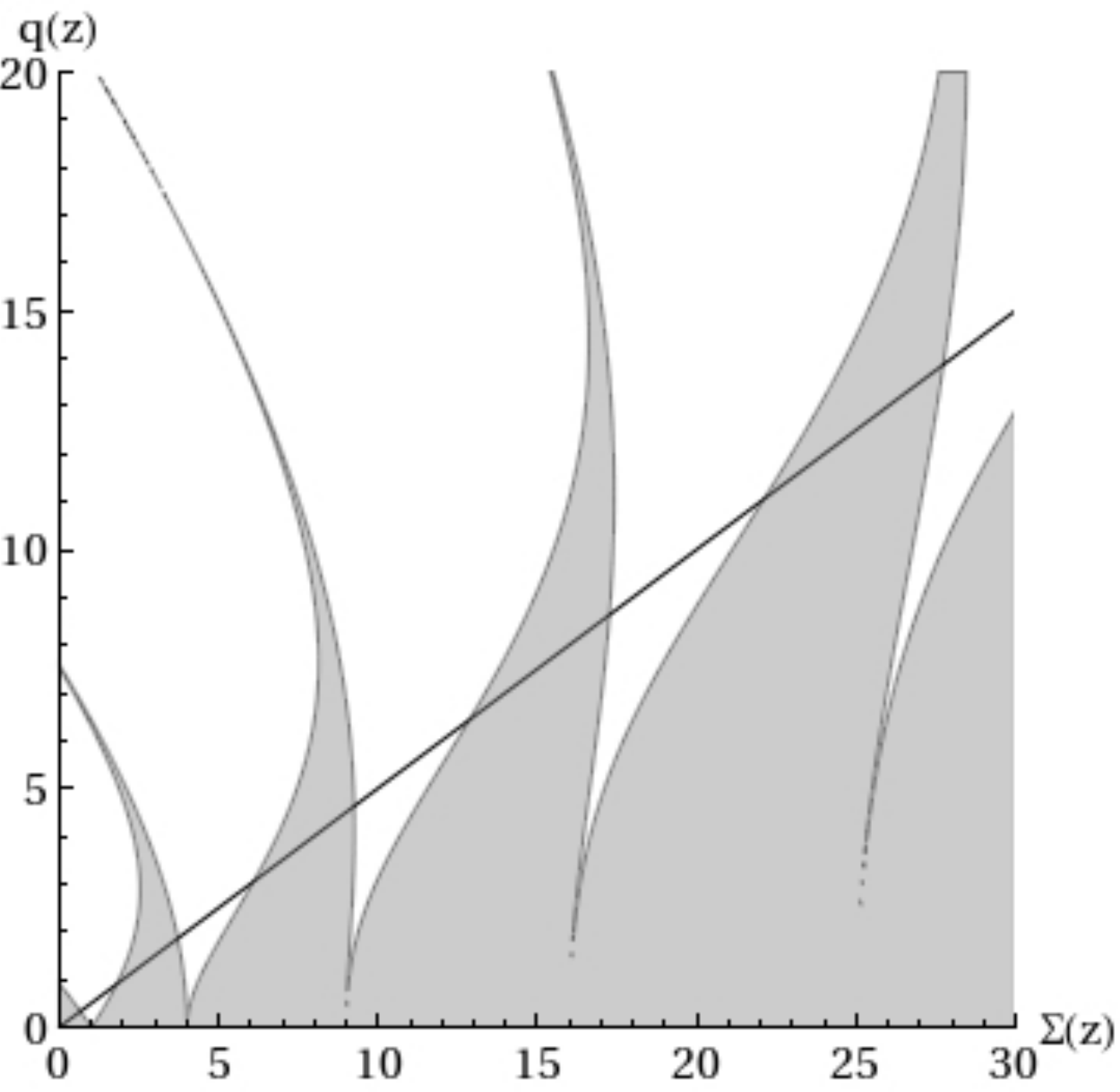}
\caption{\footnotesize{Instability chart of the Mathieu equation. Shaded regions show the stable bands; white regions show the resonance bands with exponentially growing solutions. The solid line shows $\Sigma = 2q$. For $k=0$, the solution moves towards the origin following this line very closely. The starting position and speed at which it moves depend on $m$ and $\sigma_*$. Modes with $k>0$ follow a similar evolution, but along a shallower line. Thus, modes with higher $k$ spend less time in the instability bands, leading to a weaker resonance.
}}
\label{fig:res2}
\end{center}
\end{figure}

Depending on the parameters $\Sigma$ and $q$, the solutions of the Mathieu equation are either oscillatory or exponentially growing. The growing solutions correspond to rapid, resonant energy transfer from the curvaton to the U(1) gauge field. The parameter values for which this happens form instability bands, as shown in \fig{fig:res2} (shaded regions are stable). In our case, both parameters $q(z)$ and $\Sigma_k(z)$ decrease with time, so that if they are initially large, they move through instability bands, until they leave the last instability band when $q\approx 1$, and the resonance ends. The resonance can also end if backreaction becomes so significant that the linear approximation~(\ref{eomA}) fails. The trajectory for $k=0$ modes is shown in \fig{fig:res2}. Both the speed at which the solution moves to small values of the parameters, and the initial value depend on $m$ and $\sigma_*$. The trajectory for modes with $k > 0$ is shallower, and thus modes with large $k$ do not spend enough time in the
instability bands in order to produce a resonance.

\begin{figure}[!htb]
\begin{center}
\includegraphics[width=0.6\textwidth]{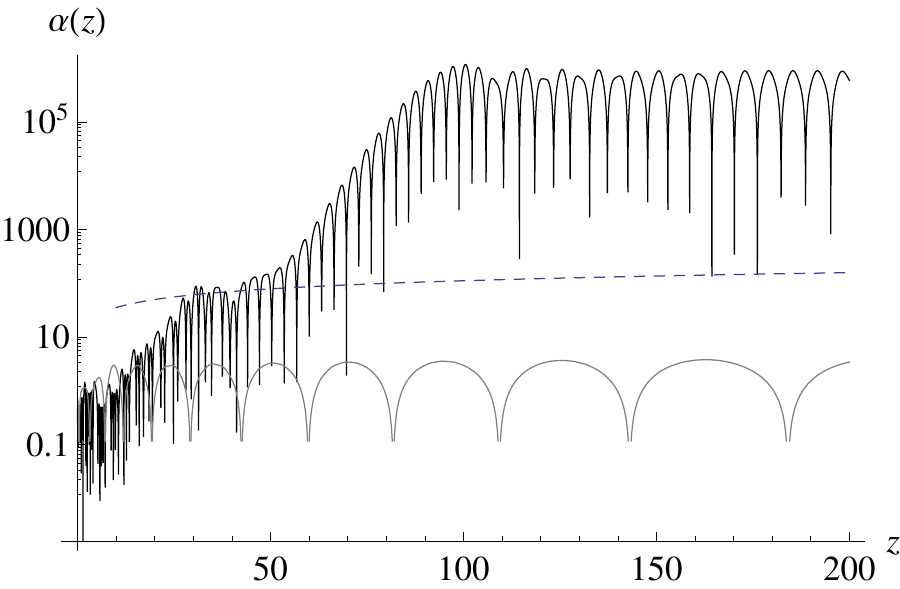}
\caption{\footnotesize{Amplification factor $\alpha$ as a function of $z$ for $k = 0$, $\sigma_* = 8 H_*$ and $m = 0.1 H_*$ (black, upper curve), $m = 0.7 H_*$ (grey, lower curve). The dashed line shows the nonlinearity condition (\ref{nonline}). For $m = 0.1H_*$ a huge amplification is seen and the resonance is clearly nonlinear. For $m = 0.7H_*$, the amplification is much less dramatic and the evolution is expected to remain linear.}}
\label{fig:A2}
\end{center}
\end{figure}

\begin{figure}[!ht]
\begin{center}
\includegraphics[width=0.45\textwidth, angle=270]{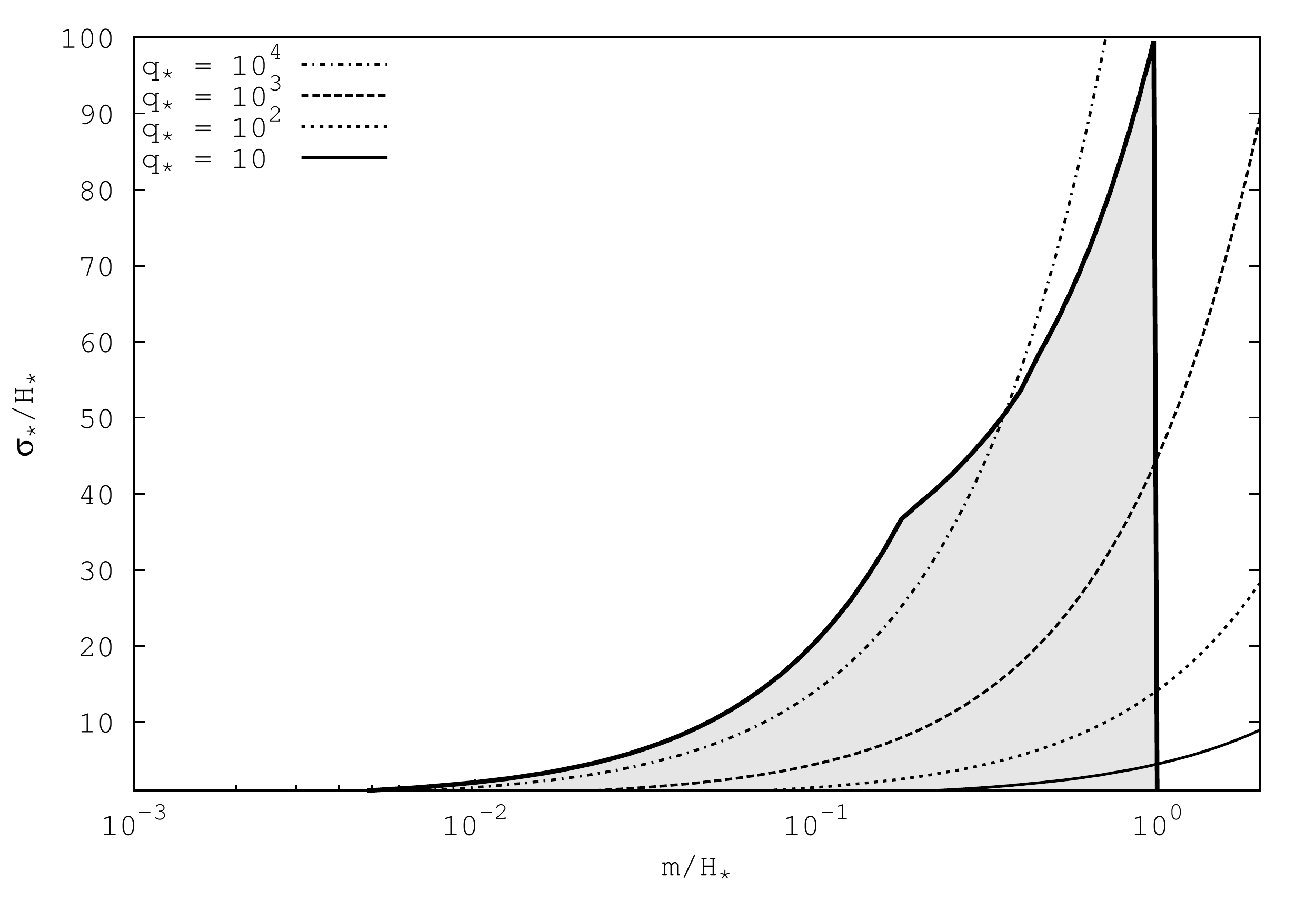}
\caption{\footnotesize{Initial resonance parameter $q_*$ in the allowed parameter space (shaded). The parameter is largest for large $\sigma_*$, small $m$. A stronger, non-linear resonance is expected for larger $q_*$.}}
\label{fig:qstar}
\end{center}
\end{figure}

In order to estimate the amount of energy transferred from the curvaton fields, we solve \eq{Mathieu} numerically, finding the amplification factor $\alpha(z)=A_{i}(z,\mathbf{k})/A_{i}(0,\mathbf{k})$. This is shown in \fig{fig:A2} as a function of the time variable $z$ for two sets of parameters.
If the amplification factor becomes large, then backreaction can no longer be ignored. We estimate that
this happens when
\be{nonlin}
\frac{g'^2\langle \mathbf{A}^2\rangle}{a(t)^2}\approx m^2.
\ee
To approximate this, we assume that modes with $k\lesssim k_{\rm res}\approx \sqrt{g' m \sigma_*}$
are all amplified by a factor $\alpha$, and we find
\be{A2amp}
\frac{g'^2\langle \mathbf{A}^2\rangle}{a(t)^2}\approx \frac{g'^2}{8\pi^2}\left(\alpha^2-1\right)
\frac{k_{\rm res}^2}{H_*t},
\ee
so that the resonance is non-linear if
\be{nonline}
\alpha(z)^2\gtrsim 10^{3} \frac{z}{\sigma_*/H_*} +1.
\ee
In many cases, only a reasonably modest amplification factor $\alpha$ is required in order for the resonance to become non-linear. If the initial value $q_*\approx (g'\sigma_*/m)^2/2$ of the resonance parameter $q$ is large, the parameters $q$ and $\Sigma$ move slowly through a large number of resonance bands. Therefore the amplification factor $\alpha$ becomes exponentially large, and we expect that the dynamics become non-linear. In this case a full numerical study is necessary to determine the dynamics. If, on the other hand,  $q_* \sim 1$, then the system moves quickly through the resonance bands, and we do not expect significant non-linear effects. \fig{fig:qstar} shows this initial value of the resonance parameter.

The value of $z$ when the resonance either ends or becomes non-linear is $z_{\rm res} \lesssim 1000$. It has been shown that for a non-linear resonance, non-relativistic curvaton particles are likely to remain after the resonance has completed \cite{Chambers:2009ki}. As a very simple estimate we assume that a fraction $f = 0.5$ of the condensate's energy density is transformed to relativistic photons, and half remains as either condensate or non-relativistic curvaton particles. This assumption allows us to use the standard expression (\ref{zeta}) for the curvature perturbation. However, simulations in a scalar model (see the bottom panels in Fig.~3 of Ref.~\cite{Chambers:2009ki}) show that the fraction $f$ is actually highly dependent on the curvaton field value, and as discussed in Section~\ref{over}, this dependence modifies the predicted curvature perturbation significantly. For fully reliable predictions, a non-linear field theory calculation is therefore required.

The fraction of the original curvaton which does \emph{not} decay resonantly can both interact with the newly-produced thermal bath and decay perturbatively (to fermions). The final $\zeta$ will include contributions from each process.
There is also a direct contribution to the curvature perturbation generated during the resonance. Its magnitude is determined by
\be{r_res}
r(t_{\rm res}) =0.74  \frac{f \sigma_*^2 (m t_{\rm res})^{1/2}}{M_p^2}.
\ee
Even by using maximum values $f = 1$, $\sigma_* = 100 H_*$ and $z_{\rm res} = 1000$, we obtain $r(t_{\rm res}) \approx 10^{-12} (H_*/10^{10}\GeV)^2$, which gives $\zeta_{\rm res} \sim 10^{-14}$, far below the observed value of $10^{-5}$. Thus, we conclude that the direct contribution to the observed $\zeta$ from the resonant decay is negligible.

\subsection{Subsequent evolution}
Once the resonance has begun, a thermal bath of photons is created. The remaining condensate can then interact with this thermal bath. The temperature is given by
\bea{T_th}
T &\simeq & 0.4  \rho_{\rm th}^{1/4} \nl
& \approx & 0.4 f^{1/4} \, (m t_{\rm res})^{1/8}\, \sqb{\sigma_*}{t}
\eea
and the effective rate of thermal interactions (using \eq{Gamma_thermal}) is

\be{Gam_th}
\Gamma_{\rm th}(t\geq t_{\rm res}) = 0.011 g'^2 f^{1/4} \, (m t_{\rm res})^{1/8}\, \sqb{\sigma_*}{t}.
\ee
The thermal interactions will eventually become important when $1/H - 1/H_{\rm res} \approx 1/\Gamma_{\rm th}$,
giving
\bea{j_th}
t_{\rm th} &=& 8.2\times 10^3 \frac{1}{g'^4 \sigma_* f^{1/2} (m t_{\rm res})^{1/4}} + t_{\rm res}.
\eea
 Evolution after the thermal interactions strongly depends on $T/m$ and we now discuss the two limiting cases.

If $T \ll m$, then the condensate decays into non-relativistic curvaton particles. Because the energy density $\rho_\sigma$ is still proportional to $1/a^3$, this does not affect the curvature perturbation $\zeta$.
Therefore, it is determined by the perturbative decay, which takes place at $t_{\rm pert} \sim 1/\Gamma$, where $\Gamma$ is the perturbative decay rate. The curvature perturbation $\zeta$ is then given by \eq{zeta}, using $r(t_{\rm pert})$.

If $T \gg m$, the thermalised curvatons are ultrarelativistic and
the evolution depends on whether chemical equilibrium is reached. This is determined by $t_{\rm chem} = 1/\Gamma_{\rm chem}$
(\eq{Gamma_chem}), which should be compared to both the timescale of the expansion ($1/H$) and the time until the relativistic curvaton freezes out again ($t_{\rm freeze}$). Freeze out will occur when $T \sim m$, i.e.
\be{z_freeze}
t_{\rm freeze} = \frac{0.14 f^{1/2} (m t_{\rm res})^{1/4} \sigma_*}{m^2}.
\ee
Thus, chemical equilibrium occurs if
\be{chem_cond}
t_{\rm chem} \ll t_{\rm th}\;\;\mbox{and}\;\;t_{\rm chem} \ll t_{\rm freeze} - t_{\rm th}.
\ee

In the case that chemical equilibrium occurs, the curvature perturbation $\zeta$ becomes adiabatic, and therefore its final value is determined by the curvaton energy fraction at the time of equilibration $r_{\rm th}$. Similarly to \eq{r_res}, we obtain
\be{r_th}
r(t_{\rm th}) =0.74  \frac{(1-f) \sigma_*^2 (m t_{\rm th})^{1/2}}{M_p^2}.
\ee
Using maximum values $f = 0.1$, $\sigma_* = 100 H_*$, $m= H_*$ and $z_{\rm res} = 100$, we find $t_{\rm th} \approx 10^{5}$ and thus that $r_{\rm th} \approx 10^{-11} (H_*/10^{10}\GeV)^2$. This gives $\zeta_{\rm th} \sim 10^{-13}$, again far below the observed value of $10^{-5}$. Thus, we conclude that if the curvaton reaches chemical equilibrium with the thermal bath, then the contribution to $\zeta$ is negligible.

If, on the other hand, the curvaton does not reach chemical equilibrium, it will eventually become non-relativistic again. The ultra-relativistic period reduces the amplitude of the curvature perturbation $\zeta$. In order to obtain the observed amplitude, the perturbative decay must then be delayed compared with the non-relativistic case. In this case, $\rho_{\sigma}$ is given by
\be{rho_sig_freeze}
\rho_{\sigma} =  0.74 \frac{m^{1/2}\sigma_*^2}{t^{3/2}} \frac{t_{\rm th}}{t_{\rm freeze}}.
\ee
The calculation of $\zeta$ then follows exactly the same procedure as for the non-relativistic case. The only difference is that a somewhat smaller $\Gamma$ will be required to obtain sufficient $\zeta$, and in some cases this value of $\Gamma$ could be ruled out by the BBN limit.

\section{Observational constraints}
\label{results}

We explore the parameter space numerically, focusing on the observables $\zeta$ (curvature perturbation), $f_{\rm NL}$ (non-Gaussianity) and $n$ (CMB spectral index) in turn.

\subsection{Curvature perturbation}
\begin{figure}[!htb]\begin{centering}
    \includegraphics[width=0.5\textwidth, angle=270]{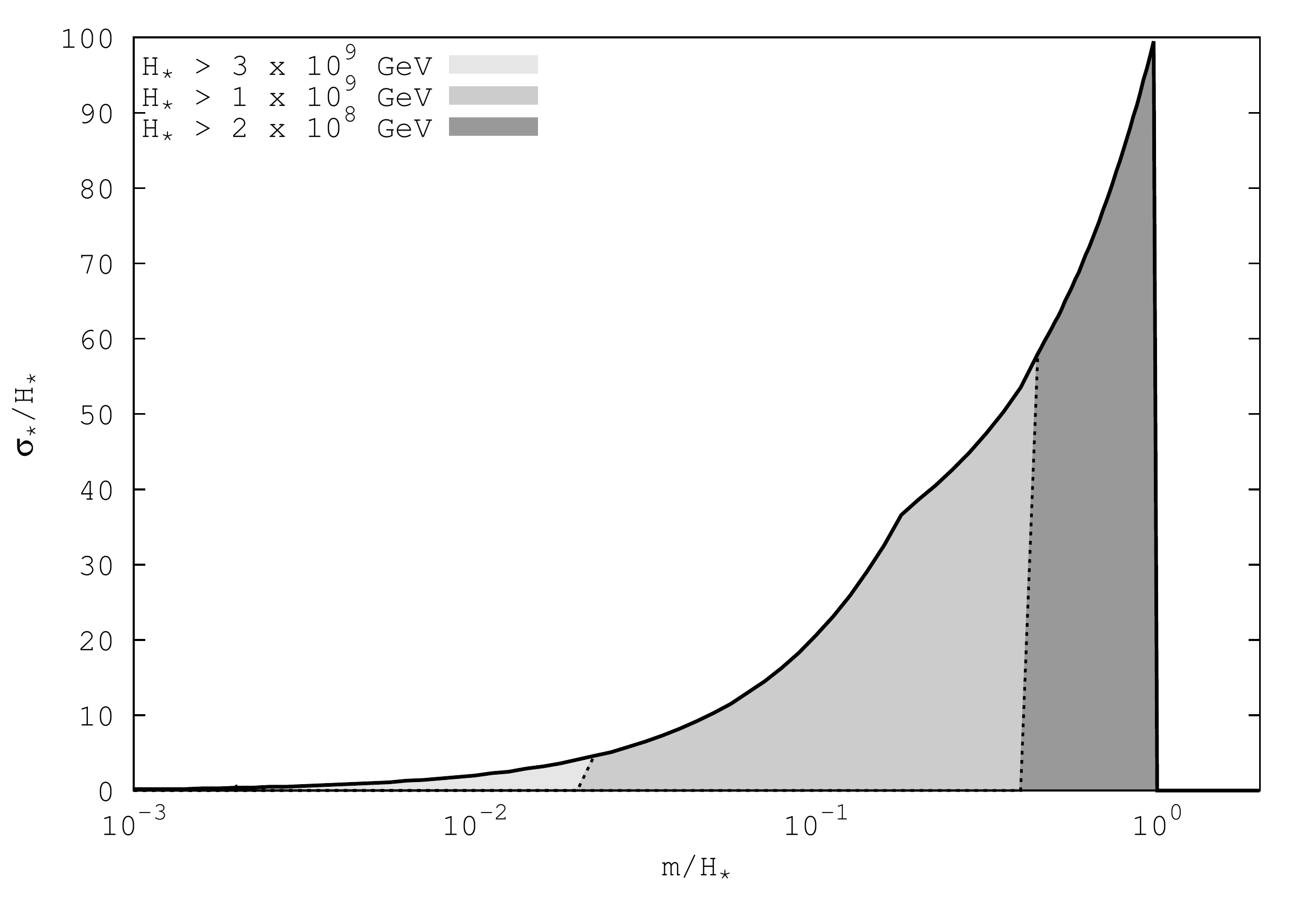}
    \caption{\footnotesize{Allowed region for a viable curvaton model which produces $\zeta = 10^{-5}$. For $H_* \gtrsim 3\times 10^{9}$, the requirement  is satisfied in the entire parameter space (all three shades of grey). The size of the allowed region reduces as $H_*$ reduces. For $H_* = 10^9\GeV$, the allowed region comprises of the two darkest grey regions; for $H_* = 2\times 10^8\GeV$ it is only the dark grey region. For $H_* \lesssim 10^{8}\GeV$ there is no allowed parameter space.} \label{pspace_zeta}}\end{centering}
\end{figure}

In general, the predicted amplitude (\ref{zeta0}) of the curvature perturbation $\zeta$ depends on the perturbative decay rate $\Gamma$, because it sets the value of $t_{\rm dec}$. Where possible, we fix $\Gamma$ in order to obtain $\zeta = 10^{-5}$, however there is a constraint on $\Gamma$. The curvaton must decay before BBN, which means that extremely small values of $\Gamma$ are not allowed. This restricts the low $H_*$, low $m$ region of parameter space, because $r \propto (\sigma_*/H_*)^2 (H_*/M_p)^2 (m/\Gamma)^{1/2}$. Thus low $m$, low $H_*$ gives low $r$. For $H_* \lesssim 10^{8}\GeV$ the requirement $\zeta = 10^{-5}$ means that there is no available parameter space left. For $10^8 \GeV \lesssim H_* \lesssim 10^9 \GeV$, the parameter space is reduced (\fig{pspace_zeta}). For $H_* \gtrsim 10^9\GeV$, there is no effect on the parameter space.

\subsection{Non-Gaussianity}
The observational limits on the non-Gaussianity of the curvature perturbation impose further constraints on the parameter space of our model. To obtain a rough estimate, we ignore the effect of the parametric resonance and use the standard expression (\ref{fnl}) for $f_{\rm NL}$.
In the parameter space with the correct $\zeta$, we can use \eq{zeta} to find $r_{\rm dec}$, giving
\be{r_dec_fnl}
r_{\rm dec} = 3\pi \zeta \frac{\sigma_*}{H_*}.
\ee
Using $\sigma_*/H_* \lesssim 100$ gives,
\be{f_NL_range}
f_{\rm NL} \gtrsim 130.
\ee
The 95\% confidence level WMAP exclusion is $-10 < f_{\rm NL} < 74$ \cite{Komatsu:2010fb}.
Being within this range would require $\sigma_* \gtrsim 180 H_*$, so this would appear to rule out the model.
However, this conclusion is not warranted because, as discussed in Section~\ref{over}, the parametric resonance modifies the predictions significantly. Thus, numerical lattice simulations are necessary.

\subsection{Spectral index}
Another observable parameter that could be used to constrain the model is the spectral index $n$. Unlike $\zeta$ and $f_{\rm NL}$, it depends on the specific model of inflation, and therefore the results are less generally applicable. As an example, we consider the simple monomial potential
\be{pot_inf}
V(\phi) = \lambda_\phi \phi^4,
\ee
This is an obvious choice in our case, because we require a non-thermal inflaton background after inflation that behaves like radiation. Requiring $N$ e-foldings of inflation gives $\phi_N^2~\approx~8N M_p^2$. The spectral index for the curvature perturbation is
\bea{n_curv}
n &=& 1 - 2\epsilon_{\rm inf} + 2\eta_{\rm curv} \nl
 &\simeq & 1- \frac{2}{N} + \frac{2 m^2}{3H_*^2}.
\eea
The current observational limits (two-sigma) are $0.939 < n < 0.995$ \cite{Komatsu:2010fb}. For $N = 50$ and $N=60$, $n$ is within the WMAP limits provided $m \lesssim 0.2H_*$. This would rule out the large $m$ region of the parameter space, but we stress that this is dependent on the model of inflation.

\subsection{Cosmic strings}
An additional interesting physical effect could occur because the curvaton field effectively has a non-zero value within any Hubble volume, therefore breaking the Standard Model U(1) symmetry spontaneously. For topological reasons there will be curves in space where the curvaton field vanishes, in very much the same way as in a cosmic string. At the end of inflation, the curvaton fields starts to oscillate, and these strings dissolve, but because the field value is zero at the string locations, they can potentially influence the curvature perturbation on cosmological scales. Furthermore, just like ordinary cosmic strings, these strings carry one quantum of magnetic flux $\Phi_0=2\pi/g'$, which can also have an effect on the reheating dynamics locally. An investigation of these effects is currently underway, using lattice simulations.

\section{Conclusions}
\label{conclusions}
We have demonstrated that it is possible to have a consistent model of the early Universe in which a scalar field charged under the Standard Model $U(1)$ weak hypercharge plays the role of the curvaton, generating a nearly scale-invariant spectrum of curvature perturbations with the observed amplitude. Besides curvature perturbations, the curvaton charge gives several other potentially observable effects. In principle, field fluctuations during inflation generate significant electric charge fluctuations on superhorizon scales, but we found that these are suppressed by the screening provided by Schwinger pair creation of other charged particles such as electron-positron pairs during inflation.

The standard calculation for this model predicts relatively a high Hubble rate during inflation, $H_* \gtrsim 10^8\GeV$, and significant non-Gaussianity ($f_{\rm NL} \gtrsim 130$), which would rule it out at 95\% confidence level by WMAP.
However, this ignores a period of parametric resonance between the curvaton and photon fields, whose effect on the curvature perturbation can only be calculated with numerical lattice field simulations. There is also freedom to relax some of the constraints in \sect{curvaton}, such as allowing a meta-stable vacuum or permitting a quartic term to initially dominate the curvaton potential.

Inflaton dynamics after inflation are also strongly constrained to avoid thermal photons destroying the curvaton condensate. In practice, this means that inflaton reheating should be substantially delayed, and that the inflaton should oscillate in a quartic potential after inflation (the inflaton should behave like radiation so that the relative magnitude of the curvaton can grow).

 The existence of an extra U(1) charged field, with a non-zero value, could also play a significant role in the electroweak phase transition. However, investigating these possibilities is beyond the scope of this paper, and we simply raise them here as potential
directions for future research.

In summary, this paper presents an interesting model that concretely links the dynamics of the Standard Model of particle physics with the very early Universe. In the era of new data from both LHC and Planck, the mechanism presented in this paper deserves further investigation.

\section*{Acknowledgements}
The authors would like to acknowledge the MSc dissertation of Hans Wiermans, which provided the starting point for this work. MD is supported by the Magnus Ehrnrooth Foundation, RL by the Academy of Finland grant 218322, and AR by the STFC consolidated grant ST/J000353/1. This research was also supported by the Royal Society International Joint Project JP100273.


\begin{thebibliography}{99}

\bibitem{Linde:1996gt}
  A.~D.~Linde and V.~F.~Mukhanov,
  Phys.\ Rev.\ D {\bf 56} (1997) 535
  [astro-ph/9610219].

\bibitem{Enqvist:2001zp}
  K.~Enqvist, M.~S.~Sloth,
  Nucl.\ Phys.\  {\bf B626 } (2002)  395-409.
  [hep-ph/0109214].

\bibitem{Lyth:2001nq}
  D.~H.~Lyth, D.~Wands,
  Phys.\ Lett.\  {\bf B524 } (2002)  5-14.
  [hep-ph/0110002].

\bibitem{Moroi:2001ct}
  T.~Moroi, T.~Takahashi,
  Phys.\ Lett.\  {\bf B522 } (2001)  215-221.
  [hep-ph/0110096].

\bibitem{Chambers:2009ki}
  A.~Chambers, S.~Nurmi, A.~Rajantie,
  JCAP {\bf 1001 } (2010)  012.
  [arXiv:0909.4535 [astro-ph.CO]].

\bibitem{Enqvist:2008be}
  K.~Enqvist, S.~Nurmi and G.~I.~Rigopoulos,
  JCAP {\bf 0810} (2008) 013
  [arXiv:0807.0382 [astro-ph]].

\bibitem{pdg}
J.~Beringer et al. (Particle Data Group), Phys.\ Rev.\ D {\bf 86} (2012) 010001.

\bibitem{Enqvist:2009zf}
  K.~Enqvist, S.~Nurmi, G.~Rigopoulos, O.~Taanila and T.~Takahashi,
  JCAP {\bf 0911} (2009) 003
  [arXiv:0906.3126 [astro-ph.CO]].

\bibitem{Enqvist:2009ww}
  K.~Enqvist, S.~Nurmi, O.~Taanila and T.~Takahashi,
  JCAP {\bf 1004} (2010) 009
  [arXiv:0912.4657 [astro-ph.CO]].

\bibitem{Chambers:2007se}
  A.~Chambers and A.~Rajantie,
  Phys.\ Rev.\ Lett.\  {\bf 100} (2008) 041302
   [Erratum-ibid.\  {\bf 101} (2008) 149903]
  [arXiv:0710.4133 [astro-ph]].

\bibitem{Chambers:2008gu}
  A.~Chambers and A.~Rajantie,
  JCAP {\bf 0808} (2008) 002
  [arXiv:0805.4795 [astro-ph]].

\bibitem{Starobinsky:1994bd}
  A.~A.~Starobinsky and J.~Yokoyama,
  Phys.\ Rev.\ D {\bf 50} (1994) 6357
  [astro-ph/9407016].

\bibitem{Enqvist:2012xn}
  K.~Enqvist, R.~N.~Lerner, O.~Taanila and A.~Tranberg,
  arXiv:1205.5446 [astro-ph.CO].

\bibitem{Caprini:2003gz}
  C.~Caprini, S.~Biller and P.~G.~Ferreira,
  JCAP {\bf 0502} (2005) 006
  [hep-ph/0310066].

\bibitem{Schwinger:1951nm}
  J.~S.~Schwinger,
  Phys.\ Rev.\  {\bf 82} (1951) 664.

\bibitem{Enqvist:2011jf}
  K.~Enqvist, R.~N.~Lerner and O.~Taanila,
  JCAP {\bf 1112} (2011) 016
  [arXiv:1105.0498 [astro-ph.CO]].

\bibitem{Coleman:1973jx}
  S.~R.~Coleman, E.~J.~Weinberg,
  Phys.\ Rev.\  {\bf D7 } (1973)  1888-1910.

\bibitem{Thoma:1996ag}
  M.~H.~Thoma and C.~T.~Traxler,
  Phys.\ Lett.\ B {\bf 378} (1996) 233
  [hep-ph/9601254].

\bibitem{Rajantie:2000fd}
  A.~Rajantie and E.~J.~Copeland,
  Phys.\ Rev.\ Lett.\  {\bf 85} (2000) 916
  [hep-ph/0003025].

\bibitem{Kofman:1994rk}
  L.~Kofman, A.~D.~Linde and A.~A.~Starobinsky,
  Phys.\ Rev.\ Lett.\  {\bf 73} (1994) 3195
  [hep-th/9405187].

\bibitem{Traschen:1990sw}
  J.~H.~Traschen and R.~H.~Brandenberger,
  Phys.\ Rev.\ D {\bf 42} (1990) 2491.

\bibitem{Komatsu:2010fb}
  E.~Komatsu {\it et al.} [ WMAP Collaboration ],
  Astrophys.\ J.\ Suppl.\  {\bf 192 } (2011)  18.
  [arXiv:1001.4538 [astro-ph.CO]].





\end{thebibliography}
\end{document}